\input{aipcheck}

\documentclass[
    ,final            
  ,numberedheadings 
  ]
  {aipproc}

\usepackage{amssymb,amsmath}
\usepackage{colortbl}

\layoutstyle{6x9}

\newcommand{\e}{{\rm e}}
\newcommand{\ii}{{\rm i}}
\newcommand{\dd}{{\rm d}}
\newcommand{\eqn}[1]{(\ref{#1})}

\def\appendix#1{\addtocounter{section}{1}\setcounter{equation}{0}
\renewcommand{\thesection}{\Alph{section}}
\section*{
\thesection\protect\indent \parbox[t]{11.715cm} {#1}}
\addcontentsline{toc}{section}{Appendix\thesection\ \ \ #1} }
\newcommand{\complex}{{\mathbb C}} 
\newcommand{\real}{{\mathbb R}} 

\newcommand{\Tr}[1]{\:{\rm Tr}\,#1}
\newcommand{\be}{\begin{equation}}
\newcommand{\ee}{\end{equation}}
\newcommand{\beq}{\begin{equation}}
\newcommand{\eeq}{\end{equation}}
\newcommand{\bea}{\begin{eqnarray}}
\newcommand{\eea}{\end{eqnarray}}
\def\beqa{\begin{eqnarray}}
\def\eeqa{\end{eqnarray}}
\def\nn{\nonumber}
\newcommand{\del}{\partial}

\newcommand{\eq}{\begin{equation}}
\newcommand{\eqa}{\begin{eqnarray}}
\newcommand{\en}{\end{equation}}
\newcommand{\ena}{\end{eqnarray}}

\newcommand{\X}{\mathfrak X}

 \def\one{\mbox{1 \kern-.59em {\rm l}}}

\definecolor{light}{gray}{.80}
\definecolor{lightt}{gray}{.90}
\definecolor{dark}{gray}{.40}
\definecolor{GreenYellow}{cmyk}{0.15,0,0.69,0}

\begin{document}

\begin{flushright}

\baselineskip=12pt
ICCUB-10-007
\\ \hfill{ }
\end{flushright}

\title{Matrix Models,
Emergent Spacetime\\ and Symmetry Breaking\footnote{Proceedings
of Fedele Lizzi's talk at the XXV Max Born
Symposium: {\sl Panck Scale}, Wroclaw 29~June-3~July 2009.}}


\author{Harald Grosse}{
  address={Department of Physics\\University of Vienna\\Boltzamnngasse5\\A-190 Vienna, Austria}
}

\author{Fedele Lizzi}{
  address={Dipartimento di Scienze Fisiche, Universit\`{a}
di Napoli {\sl Federico II}\\ and  INFN, Sezione di Napoli\\
 Monte S.~Angelo, Via Cintia, 80126 Napoli, Italy}
  ,altaddress={ Institut de Ci\'encies del Cosmos,
Universitat de Barcelona,\\
Mart\'i i Franq\`ues 1, Barcelona, Catalonia, Spain} }

\author{Harold Steinacker}{
  address={Department of Physics\\University of Vienna\\Boltzamnngasse5\\A-190 Vienna,
  Austria}
   }

\begin{abstract}
We discuss how a matrix model recently shown to describe
emergent gravity may contain extra degrees of freedom which
reproduce some characteristics of the standard model, in
particular the breaking of symmetries and the correct quantum numbers of fermions.
\end{abstract}

\maketitle


\section{Introduction}

One of the recurrent themes of this conference is the fact that at
the Planck scale ordinary geometry may no longer be valid, and that
a quantized theory of gravity could be bases on this new, more
general, geometry. The guiding analogy is with the transition from
the classical to the quantum phase space. The former is described by
ordinary geometry, that made of points, lines, tangents etc. The
latter requires a \emph{noncommutative
geometry}~\cite{connes,landi,ticos} better described by operators on
an Hilbert space. In this case the coordinates and in general the
observables of phase space become noncommuting operators on an
Hilbert space. The commutator of the coordinates is a constant
(Planck's constant) and therefore this kind of noncommutative
geometry is in some sense the simplest. A noncommutative geometry of
spacetime (as opposed to phase space) is then the one described by
the commutator
 $
[x^\mu,x^\nu]=\ii\theta^{\mu\nu} $, where $\theta^{\mu\nu}$ is
\emph{constant} quantity of the order of the square of Planck's
length. This has led to the study of field theories on
noncommutative space, where the noncommutativity of the space is
implemented by the use of the Gr\"onewold-Moyal $\star$ product, for
a review see~\cite{Szaboreview}. While this is not the place to
describe successes and problems of this approach, it is clear, at
least to me, that the implementation of noncommutativity with a
constant $\theta$ is at best an approximation of a more general
theory.

In these proceedings we will describe how the matrix model introduced in~\cite{EguchiKawai, IKKT}, and which was shown in~\cite{Steinackeroriginal} to describes emergent
gravity, may lead to an understanding  of aspects of the
standard model, somehow along the line of the Connes vision of the
standard model~\cite{connes, ConnesLott, AC2M2}.

\section{Noncommutative Spacetime and the Matrix Model}
\setcounter{equation}{0}

The general programme is to describe noncommutative spacetime as a
matrix model with a simple action, inserting fermions, and have
gravity, and other physical characteristic, emerge as fluctuations
around a semiclassical vacuum. We will be very impressionistic and
concise for because of lenght restrictions, but not only, it should
be born in mind that the work present is still in progress. More
details can be found in~\cite{GLS}.

The rationale behind this approach is quite simple and can be
\emph{very} heuristically (and therefore incorrectly) be stated as
follows:

\begin{itemize}

\item A noncommutative geometry is described by a noncommutative
    algebra, deformation of the commutative algebra of function
    on some space

\item Any noncommutative ($C^*$)-algebra is represented as
    operators on some Hilbert space

\item Operators on a Hilbert space are just infinite matrices

\end{itemize}

Consider a $U(1)$ gauge theory in a space described by the
Gr\"onewold-Moyal $\star$ product. The theory is noncommutative
(also in the $U(1)$ case), due to the noncommutativity of the
product. In this sense we can talk of a noncommutative geometry,
because we substitute the commutative algebra of functions on
spacetime by the algebra obtained deforming the product as:
\be
\left.(f\star
g)(x)=\e^{\frac{\ii}{2}\theta^{\mu\nu}\del_{y_\mu}\del_{z_\nu}}f(y)g(z)\right\vert_{x=y=z}
\label{Moyalseries}
\ee
so that the noncommutativity of the coordinates is encoded on the
commutation relation
\be
x^\mu\star x^\nu-x^\nu\star
x^\mu=[x^\mu,x^\nu]_\star=\ii\theta^{\mu\nu}
\ee
and derivation become inner, i.e.\ they can be expressed by a
commutator:
\be
\frac\del{\del x^\mu} f= \ii(\theta)^{-1}_{\mu\nu}[x^\nu,f]
\label{derivcomm}
\ee
The field strength  of the $U(1)$ theory is modified by
\be
F_{\mu\nu}=\del_\mu A_\nu - \del_\nu A_\mu - \ii[A_\mu,A_\nu]_\star
\ee
Consider as usual the action to be the square of the curvature:
\be
S=-\frac14\int \dd x \, F_{\mu\nu}\star F^{\mu\nu}
\ee
The theory is invariant for the gauge transformation $F\to U\star
F\star U^\dagger$ where $U$ is $\star$-unitary: $U\star
U^\dagger=1$. A nonabelian (and noncommutative) Yang-Mills gauge
theory is obtained considering $A_\mu=A_\mu^\alpha\lambda^\alpha$
for $\lambda^\alpha$ generators of $U(n)$.

Because of the connection between commutator with the coordinates
and derivatives~\eqn{derivcomm}, one is led~\cite{MSSW} to the
definition of \emph{covariant coordinates}
\be
X^\mu=x^\mu + \theta^{\mu\nu} A_\nu
\ee
and consequently
\be
D_\mu f =\ii\theta^{-1}_{\mu\nu}[X^\mu,f]_\star=\del_\mu f - \ii
[f,A_\mu]_\star
\ee
and therefore we have
\be
F^{\mu\nu}=[D^\mu,D^\nu]_\star=[X^\mu,X^\nu]_\star + \theta^{\mu\nu}
\ee
The constant $\theta$ can be reabsorbed by a field redefinition and
the action is the square of this quantity, integrated over
spacetime.

The objects we have defined are elements of a noncommutative algebra
and we can always represent them as operators on a Hilbert space, in
this case the integral becomes a trace and this suggests the use of
the matrix following action
\be
S=-\frac1{4g}\Tr
[X^\mu,X^\nu][X^{\mu'},X^{\nu'}]g_{\mu\mu'}g_{\nu\nu'}
\label{bosonicaction}
\ee
where the $X$'s are operators (matrices) and the metric
$g_{\mu\mu'}$ is the \emph{flat} Minkowski (or Euclidean) metric.
The important characteristic of this action is that gravity
\emph{emerges} naturally from it~\cite{Steinackeroriginal}.

The equations of motion corresponding to the
action~\eqn{bosonicaction} are
\be
[X^\mu,[X^\nu,X^{\mu'}]]g_{\mu\mu'}=0
\ee
A possible vacuum (which we may call the $U(1)$ Gr\"onewold-Moyal
vacuum) is given by a set of matrices $X_0$ such that
\be
[X_0^\mu,X_0^\nu]=\ii\theta^{\mu\nu}
\ee
with $\theta$ constant. This corresponds to the Gr\"onewold-Moyal
case and in this case the vacuum is just the deformation of
spacetime described earlier. Now let fluctuate these ``coordinates''
and consider $X^\mu=X_0^\mu+A^\mu$ so that
$[X^\mu,X^\nu]=\ii\theta(X)$ and we are considering a nonconstant
noncommutativity.

Gravity emerges as nontrivial curvature considering the coupling
with a scalar field $\Sigma$. The (free) action is then
\be
\Tr [X^\mu,\Sigma][X^\nu,\Sigma]g_{\mu\nu}\sim\int\dd x
(D_{\mu'}\Sigma)
(D_{\nu'}\Sigma)\theta^{\mu\mu'}\theta^{\nu\nu'}g_{\mu\nu}= \int\dd
x (D_{\mu}\Sigma)(D_{\nu}\Sigma)G^{\mu\nu}
\ee
which describes the coupling of the scalar field to a curved
background defined by the new (non flat) metric
$G^{\mu\nu}(x)=\theta^{\mu\mu'}\theta^{\nu\nu'}g_{\mu'\nu'}$. A
curved background emerges in an effective way and the gravitational
action is recovered as an effective action at one loop, for details
see~\cite{Steinackeroriginal}.

\section{Alternative Vacua and Nonabelian Symmetry}
\setcounter{equation}{0}

An alternative vacuum, still solution of the equations of motion, is
\be
\bar X^\mu_0=X_0^\mu\otimes\one_{n}
\ee
In this case the Moyal-Weyl limit is given by matrix valued
functions on $\real^D_\theta$ and the gauge symmetry is given by
unitary elements of the algebra of $n\times n$ matrices of functions
of the $X_0$. This theory therefore has a noncommutative $U(n)$
gauge symmetry because in the semiclassical limit it corresponds to
a nonabelian gauge theory. However the $U(1)$ degree of freedom is
the one which couples gravitationally for the emergent gravity,
hence the gauge theory in this case is a nonabelian $SU(n)$ theory.
In fact consider the fluctuations of $\bar x_0$ to be
\be
\bar X=\bar X_0+A_0+A_\alpha\lambda_\alpha
\ee
Where in the fluctuations we have separated the traceless generators
of $SU(n)$ from the trace part ($A_0$). The $U(1)$ trace part of the
fluctuation gives rise to the gravitational coupling, while the
remaining $A_\alpha$ describe an $SU(n)$ gauge theory.

Once we are convinced that we can reproduce a $SU(n)$ gauge theory
the next objective is to reproduce the standard model, including the
symmetry breaking, or at least some gauge theory which closely
resembles resembles it. Therefore the game becomes to find a
noncommutative geometry, described by a matrix model, with a bosonic
action like (even if possibly not identical to)~\eqn{bosonicaction},
with the insertion of fermions transforming properly under the gauge
group and a symmetry breaking mechanism. We should not be shy of
making as many assumptions as are needed. The game is not to find
the standard model, but rather to find a noncommutative geometry
which ``fits'' it. The programme in this sense is similar to the one
started by Alain Connes and collaborators~\cite{connes, ConnesLott,
AC2M2}. The main difference is the fact that in Connes' work
spacetime is still commutative, while in the present case spacetime
emerges in the limit as a Gro\"newold-Moyal space. Other differences
is the fact that we are naturally Minkowskian and that the action is
not the same. The model I will present is incomplete and can be
considered as a first approximation. Remarkably however some key
characteristics of the standard model emerge naturally, which makes
us confident that a fully viable (and predictive) model is within
reach.

Consider a another extra dimension which we call $\X^\Phi$ of the
form
\be
\X^\Phi=\left(\begin{array}{ccc}
\alpha_1\one_{2}& \  & \ \\ \ & \alpha_2\one_{2} & \ \\
\ & \ & \alpha_3\one_{3} \end{array}\right)
\ee
Where the $\alpha$'s are constants all different among themselves.
This new coordinate is still solutions of the equations of the
motion because $[X^\mu,\X^\Phi]=0$, this signifies that
$\theta^{\mu\Phi}=0$, and in turn that $G^{\Phi\Phi}=G^{\mu\Phi}=0$, whatever the
value of the metric $g_{\mu\Phi}$ and $g_{\Phi\Phi}$. Therefore the
extra coordinate is not geometric and does not correspond to
propagating degrees of freedom from the four dimensional point of
view. The new coordinate is not invariant for the transformation
\be
\X^\Phi\to U \X^\Phi U^\dagger\neq \X^\Phi
\ee
for a generic $U\in SU(7)$. It is invariant only for a subgroup of it. The traceless part of it is $SU(2)\times SU(2) \times SU(3) \times U(1) \times
U(1)$.

If we consider the gauge bosonic action  we have that the spacetime
($\mu\nu$) part of action remains unchanged, while for the $\mu\phi$
components we have that Moyal-Weyl limit is
\bea
[\bar X^\mu +A^\mu, \X^\Phi] &=& \theta^{\mu\nu} D_\nu \X^\phi
=  \theta^{\mu\nu} (\partial_\nu +  i A_\nu) \X^\phi , \nn\\
-(2\pi)^2 \Tr [\bar X^\mu,\X^\phi] [\bar X^{\nu},\X^\phi] g_{\mu\nu}
&=& \int d^4 x G^{\mu\nu} \left(\partial_\mu \X^\Phi \partial_\nu
\X^\Phi
+ [A_\mu,\X^\Phi][A_\nu,\X^\Phi]\right) \nn\\
\eea
because the mixed terms, assuming the Lorentz gauge $\partial^\mu A_\mu = 0$, vanish:
\be
\int \partial^\mu \X^\Phi [A_\mu,\X^\Phi] =
- \frac 12 \int  \X^\phi [\partial^\mu A_\mu,\X^\Phi] = 0
\ee
Since we have
that $\bar X^\mu$ and $\X^\phi$ commute, this means $\X^\Phi =
\mathrm{const}$ and the first term in the integral above vanish. We
can therefore separate the fluctuations of this extra dimension
which are a field, the
(high energy) Higgs field. In the action the first term is nothing
but the covariant derivative of it. The second term instead is
\be
[A^\mu,\X^\phi]=\left(\begin{array}{ccc}
 0 & (\alpha_2-\alpha_1) A^\mu_{12} & (\alpha_3-\alpha_1) A^\mu_{13} \\
 (\alpha_1-\alpha_2) A^\mu_{21} & 0 & (\alpha_3-\alpha_2) A^\mu_{23} \\
 (\alpha_1-\alpha_3) A^\mu_{31} & (\alpha_2-\alpha_3) A^\mu_{32} & 0
\end{array}\right)
\ee
where we consider the block form of $A^\mu$
\be
A^\mu=\left(\begin{array}{ccc}
 A^\mu_{11} & A^\mu_{12} &  A^\mu_{13} \\
 A^\mu_{21} & A^\mu_{22} &  A^\mu_{23} \\
 A^\mu_{31} &  A^\mu_{32} & A^\mu_{33}
\end{array}\right)
\ee
If we now assume that the differences $\alpha_i-\alpha_j$ are large,
say of the grand unification scale, it is easy to see that that all
non diagonal blocks of $A^\mu$ acquire large masses, thus
effectively decoupling. In this way we have reduced the symmetry to
a smaller group, resembling more the standard model.

We now assume that a this first stage of breaking has reduced the
symmetry and indicate the spacetime coordinates $X$ as
\be
X^\mu =
\begin{pmatrix}
  X_0^\mu\otimes \one_2&  &  &  \\
   & X_0^\mu& &  \\
   &  & X_0^\mu &  \\
   & &  & X_0^\mu\otimes \one_3
\end{pmatrix}
\ee
The smaller group contains the electroweak $SU(2)$ group and the
colour $SU(3)$ groups defined by:
\be
W^\mu=\begin{pmatrix}
  w^\mu &  &  &  \\
   & 0& &  \\
   &  & 0 & \\
   & &  & 0
   \end{pmatrix}
\ee
and
\be
G^\mu=\begin{pmatrix}
  0 &  &  &  \\
   & 0& &  \\
   &  & 0 & \\
   & &  & g^\mu
   \end{pmatrix}
\ee
Where $w$ and $g$ are in the adjoint representations of the
respective groups. There are also
some $U(1)$ symmetries, apart form the one generated by the trace
which as we said couples to the gravitational degrees of freedom. In
order to recognize the physically relevant factors let us now
introduce the fermions.

%
%
%

%
%

\section{Fermions and the Gauge Charge Problem}
\setcounter{equation}{0}

The action for fermions in matrix models has been introduced
in~\cite{IKKT} and discussed in this context in~\cite{KlammerSteinackerfermions} and is
\be
S_F=\Tr\bar\Psi\gamma_a[X^a,\Psi]\sim \int \dd x \bar\Psi \gamma_a
D^a\Psi
\ee
Here $\Psi$ is a $7\times 7$ whose entries are spinors, i.e.\ they
carry an extra index on which the $\gamma$ matrices act, these are
represented diagonally on the matrix space. In this note we will
choose $\Psi$ to be upper triangular and its components to be Dirac
spinors. This choice is not unique (and in the long run it may turn
out not to be the best one), for more details see~\cite{GLS}. We
accommodate all known fermions (except right handed neutrinos) as:
\be
\Psi=\left(
       \begin{array}{ccccccc}
         0 & 0 & 0 & \nu_L & {u_L}_1 & {u_L}_2 & {u_L}_2 \\
         0 & 0 & 0 & e_L & {d_L}_1 & {d_L}_2 & {d_L}_3 \\
         0 & 0 & 0 & e_R & {d_R}_1 & {d_R}_2 & {d_R}_1 \\
         0 & 0 & 0 & 0 & {u_R}_1 & {u_R}_2 & {u_R}_3 \\
         0 & 0 & 0 & 0 & 0 & 0 & 0 \\
         0 & 0 & 0 & 0 & 0 & 0 & 0 \\
         0 & 0 & 0 & 0 & 0 & 0 & 0 \\
       \end{array}
     \right)
\label{psimatrix}
\ee
Where the numerical labels (1,2,3) for quarks indicate colour. Right
handed neutrino can be inserted as diagonal elements in the $(4,4)$
position in this matrix. We do not insert them for the present
discussion because it commonly believed that their origin lies
beyond the standard model, a belief confirmed in the present model
by the fact that they will not fit in the triangular scheme. The
empty spaces may be filled by extra particles, for example there is
room for another weak doublet with the quantum numbers of the
Higgsino.

To avoid mirror fermions we choose
\be
\gamma^5\Psi=\left(
       \begin{array}{ccccccc}
         0 & 0 & 0 & \nu_L & {u_L}_1 & {u_L}_2 & {u_L}_2 \\
         0 & 0 & 0 & e_L & {d_L}_1 & {d_L}_2 & {d_L}_3 \\
         0 & 0 & 0 & -e_R & -{d_R}_1 & -{d_R}_2 & -{d_R}_1 \\
         0 & 0 & 0 & 0 & -{u_R}_1 & -{u_R}_2 & -{u_R}_3 \\
         0 & 0 & 0 & 0 & 0 & 0 & 0 \\
         0 & 0 & 0 & 0 & 0 & 0 & 0 \\
         0 & 0 & 0 & 0 & 0 & 0 & 0 \\
       \end{array}
     \right)
\ee
The fermions which appear in the matrix $\Psi$ in the present case
are chiral Dirac fermions, in the sense that they are in the
$(1/2,0)\oplus(0,1/2)$ representation of $SL(2,\complex)$, but only
the left or right component is different from zero. That is, we have
manually set to zero the ``mirror sector'' which appears naturally.
This is not new in noncommutative geometry, a similar phenomenon of fermion doubling
appears in Connes' approach~\cite{LMMS}.

The correct hypercharge, electric charge and baryon number are then
reproduced by the following generators (the constant in front of the
unit matrix is chosen in order to make them traceless)
\bea
Y &=& \begin{pmatrix}
   0_{2\times 2}  &  &  &  \\
   & - \sigma_3 & &  \\
   & &  & -\frac 13 \one_{3\times 3}
\end{pmatrix} -\frac17 \one   \label{Ygenerator}\\
Q &=& T_3 + \frac Y2 = \frac 12 \begin{pmatrix}
   \sigma_3  &  &  &  \\
   & - \sigma_3 & &  \\
   & &  & -\frac 13 \one_{3\times 3}
\end{pmatrix} -\frac27 \one  \label{Qgenerator}\\
B &=& \begin{pmatrix}
   0  &  &  &  \\
   & 0 & &  \\
   & &  & - \frac 13\one_{3\times 3}
\end{pmatrix} -\frac17 \one \label{Bgenerator}
\eea
which act in the adjoint.
Now we can write
\be
\left[Q - \frac 12 B\right] = \frac 12 \left[ \begin{matrix}
   \sigma_3 & 0 & 0\\
    0 & -\sigma_3& 0 \\
    0 & 0 & 0
\end{matrix}\right]
= \frac 12  \sigma_3 \otimes \sigma_3
\ee
where by $[\cdot]$ we indicate the $4\times 4$ matrix obtained
dropping the colour part, which plays no role in the following. It
is always understood that the remainig blocks of the $7\times 7$
matrix are zero.

The present formulation solves an oustanding problem in
noncommutative gauge theories which we will call the \emph{charge
quatization problem}~\cite{Hayakawa,CPS-JT}. In the usual
formulation of gauge theory (on commutative spaces) it is of course
possible to have different fermions transforming under different
representations of a particular gauge group. For example there are
particles with electric charge $0, \pm 1,\pm 2/3, \pm 1/3$. This is
possible because the gauge group is the tensor product of a finite
dimensional group (or algebra) times the functions on the spacetime
manifold. Different particles will belong in turn to the tensor
product of a vector belonging to the appropriate module (vector
space on which the corresponding representation of the group act)
times again the space of functions on the spacetime manifold.

In conventional \emph{noncommutative} gauge theory with a $\star$
product, the space time and the internal symmetry are intimately
intertwined, in fact the theory  is noncommutative also in the case
of $U(1)$ gauge group, the noncommutativity being given by the
product. The representations of this large gauge group are much less
than the ones of the tensor product (commutative space) case. In the
latter case it is possible to have functions transforming, in their
internal components, according to any representation of the gauge
group. In the the former case only fundamental, adjoint and singlet
cases are possible. In particular it is impossible to obtain charges
which are different from $0,\pm 1$. The problem can be solved at the
price of adding one extra $U(1)$ for each different charge, and
these extra gauge degrees of freedom have to be later spontaneously
broken.

In the matrix model formalism, although spacetime remains
noncommutative, and in the limit reproduces the one with the
Gronew\"old-Moyal~$\star$ product, the problem does not appear.
Fermions are in the adjoint representation, but the various U(1) are
generated by a single diagonal traceless matrix. Since the fermions
are off diagonal, the fermion matrix retains its form when
commuted with the generator of a $U(1)$, such as $Y$ and $Q$ in
equations~\eqn{Ygenerator} and~\eqn{Qgenerator}. For example for the
charge $Q$:
\be
[Q,\Psi]=\left(
       \begin{array}{ccccccc}
         0 & 0 & 0 & 0 & \frac23{u_L}_1 & \frac23{u_L}_2 & \frac23{u_L}_2 \\
         0 & 0 & 0 & -e_L & -\frac13{d_L}_1 & -\frac13{d_L}_2 & -\frac13{d_L}_3 \\
         0 & 0 & 0 & -e_R & -\frac13{d_R}_1 & -\frac13{d_R}_2 & -\frac13{d_R}_1 \\
         0 & 0 & 0 & 0 & \frac23{u_R}_1 & \frac23{u_R}_2 & \frac23{u_R}_3 \\
         0 & 0 & 0 & 0 & 0 & 0 & 0 \\
         0 & 0 & 0 & 0 & 0 & 0 & 0 \\
         0 & 0 & 0 & 0 & 0 & 0 & 0 \\
       \end{array}
     \right)
\ee
Which is the correct charge assignment.

Consider now the extra coordinate:
\be
\X^\varphi = \left(\begin{array}{cccccc}
  0_{2\times 2} & \varphi & 0_{2\times 1} & 0_{2\times 1} & 0_{2\times 1} & 0_{2\times 1}  \\
 \varphi^\dagger & 0 & 0 & 0 & 0& 0 \\
 0_{1\times 2} & 0 &0 &0 & 0& 0\\
 0_{1\times 2} & 0 &0 &0 & 0& 0\\
 0_{1\times 2} & 0 &0 &0 & 0& 0\\
 0_{1\times 2} & 0 &0 &0 & 0& 0\\   \end{array}\right)
\ee
Where $\varphi$ is the usual 2-component Higgs with vaccum
expectation value
\be
\langle\varphi\rangle=\begin{pmatrix} 0\\ v \end{pmatrix}
\ee

Since $[X^\mu,X^\varphi]=0$ the extra coordinate is still a solution
of the equation of motion, with $\theta^{\mu\varphi}=0$ which
implies that in the semiclassical limit $G^{\mu\varphi}=0$. In other
words this extra dimension is not dynamical.

The extra coordinate $\X^\phi$ does not commute with the generators
of the weak $SU(2)$ and with the hypercharge $Y$, but it does
commute with the electric charge~\eqn{Qgenerator}, thus breaking the
$SU(2)\times U(1)$ symmetry to $U(1)_Q$.

The computation of the Yukawa couplings is now straightforward, it
it the part of the fermionic action involving the extra coordinate
$\X^\phi$:
\be
S_Y = \Tr \bar\Psi^\dagger \gamma_\phi [\X^\phi,\Psi]
\ee
We will
also consider $\Psi$ to be eigenvalue of $\gamma^\phi$ with
eigenvalue 1.
\be
S_Y= v(d_R^\dagger d_L +u_R^\dagger u_L +e_R^\dagger e_L +
d_L^\dagger d_R +u_L^\dagger u_R + e_L^\dagger e_R)
\ee
Those are the correct mass terms, in the sense that left handed
fermions couple with the right handed ones, and the extra coordinate
has succeeded in breaking the symmetry.

\section{Conclusions}
The model presented here is just a ``feasibility study'' for the
construction of more realistic models. At the present we can manage
to reproduce some key features of it in a rather simple (and naive)
model with a pair of extra coordinates. The list of drawbacks and
physically unrealistic features of this simple model is probably too
long to be mentioned fully. It comprises the fact that at the bare
level the Yukawa couplings are the same for leptons and quarks (but
this may change under renormalization), that the unbroken gauge
group contains some extra $U(1)$ factors (probably anomalous), and the fact that there are three
generations does not appear. Some of the problems are solved
considering the internal space not being described by two single
coordinates ($\X^\Phi$ and $\X^\phi$) but having an internal
noncommutative structure, typically that of fuzzy
spheres~\cite{GLS} along the lines of~\cite{AGSZ}. As the extra dimension get more structure the
model becomes more realistic, and hopefully in a near future it will
be possible to make definite predictions. The gravitational aspect
is already being investigated from a phenomenological point of view
and in fact some cosmological predictions have already
appeared~\cite{KlammerSteinackercosmo}. We hope that we have been
able to convince that matrix models can have a semiclassical limit
describing the standard model and gravity as an emergent phenomenon.

\paragraph{Acknowledgments} One of us (FL) would like the Department of Estructura i Constituents de la Materia, and the Institut de Ci\'encies del Cosmos,
Universitat de Barcelona for hospitality. His work has been
supported in part by CUR Generalitat de Catalunya under project
2009SGR502. He would like to thank Jerzy Kowalski-Glikman,
Krzysztof Meissner, Remigiusz Durka, Anna Pachol, Michal
Szczachor and Adrian Walkus for organizing a most interesting
meeting.
The work of H. S. was supported in part by the FWF project P18657
and in part by the FWF project P21610.

\bibliographystyle{aipproc}

\begin{thebibliography}{00}

\bibitem{connes} A. Connes, {\it Noncommutative Geometry}, Academic
    Press, (1994).

\bibitem{landi} G.~Landi,
  {\it An introduction to noncommutative spaces and their
  geometry,} Springer, (1997).
  arXiv:hep-th/9701078.


\bibitem{ticos} J.~M.~Gracia-Bondia, J.~C.~Varilly and H.~Figueroa,
  {\it  Elements Of Noncommutative Geometry,}
Birkhaeuser (2001)

\bibitem{Szaboreview} R.~J.~Szabo,
  ``Quantum Field Theory on Noncommutative Spaces,''
  Phys.\ Rept.\  {\bf 378} (2003) 207
  [arXiv:hep-th/0109162].

\bibitem{ConnesLott} A.~Connes and J.~Lott,
  ``Particle Models and Noncommutative Geometry (Expanded Version),''
  Nucl.\ Phys.\ Proc.\ Suppl.\  {\bf 18B}, 29 (1991).

\bibitem{AC2M2} A.~H.~Chamseddine, A.~Connes and M.~Marcolli,
  ``Gravity and the standard model with neutrino mixing,''
  Adv.\ Theor.\ Math.\ Phys.\  {\bf 11}, 991 (2007)
  [arXiv:hep-th/0610241].

\bibitem{EguchiKawai}
T.~Eguchi and H.~Kawai,
  ``Reduction Of Dynamical Degrees Of Freedom In The Large N Gauge Theory,''
  Phys.\ Rev.\ Lett.\  {\bf 48} (1982) 1063.


\bibitem{IKKT}
N.~Ishibashi, H.~Kawai, Y.~Kitazawa and A.~Tsuchiya,
  ``A large-N reduced model as superstring,''
  Nucl.\ Phys.\  B {\bf 498} (1997) 467
  [arXiv:hep-th/9612115].



\bibitem{Steinackeroriginal} H.~Steinacker,
  ``Emergent Gravity from Noncommutative Gauge Theory,''
  JHEP {\bf 0712}, 049 (2007)
  [arXiv:0708.2426 [hep-th]].

\bibitem{GLS} H.~Grosse, F.~Lizzi and H.~Steinacker, 
``Noncommutative gauge theory and symmetry breaking in matrix
models.'' [arXiv:1001.2703 [hep-th]].


\bibitem{MSSW} J.~Madore, S.~Schraml, P.~Schupp and J.~Wess,
  ``Gauge theory on noncommutative spaces,''
  Eur.\ Phys.\ J.\  C {\bf 16} (2000) 161
  [arXiv:hep-th/0001203].

\bibitem{KlammerSteinackerfermions} D.~Klammer and H.~Steinacker,
  ``Fermions and Emergent Noncommutative Gravity,''
  JHEP {\bf 0808} (2008) 074
 {} [arXiv:0805.1157 [hep-th]].


\bibitem{LMMS} F.~Lizzi, G.~Mangano, G.~Miele and G.~Sparano,
  ``Fermion Hilbert space and fermion doubling in the noncommutative  geometry
  approach to gauge theories,''
  Phys.\ Rev.\  D {\bf 55} (1997) 6357
  [arXiv:hep-th/9610035].

\bibitem{AGSZ} P.~Aschieri, T.~Grammatikopoulos, H.~Steinacker and G.~Zoupanos,
  ``Dynamical generation of fuzzy extra dimensions, dimensional reduction and
  symmetry breaking,''
  JHEP {\bf 0609}, 026 (2006)
  [arXiv:hep-th/0606021].

\bibitem{KlammerSteinackercosmo} D.~Klammer and H.~Steinacker,
  ``Cosmological solutions of emergent noncommutative gravity,''
  Phys.\ Rev.\ Lett.\  {\bf 102} (2009) 221301
  [arXiv:0903.0986 [gr-qc]].

\bibitem{Hayakawa} M.~Hayakawa,
  ``Perturbative analysis on infrared aspects of noncommutative QED on  R**4,''
  Phys.\ Lett.\  B {\bf 478} (2000) 394
  [arXiv:hep-th/9912094].


\bibitem{CPS-JT}
M.~Chaichian, P.~Presnajder, M.~M.~Sheikh-Jabbari
    and
    A.~Tureanu,
  ``Noncommutative gauge field theories: A no-go theorem,''
  Phys.\ Lett.\  B {\bf 526}, 132 (2002)
  [arXiv:hep-th/0107037].


\end{thebibliography}

\end{document}